# Continuous Human Action Detection Based on Wearable Inertial Data

Xia Gong
*School of Music*, *Shandong University of Technology*, Zibo, China, sdlggx@sdut.edu.cn
Yan Lu
*High School Affiliated (Jia Ding) to Shanghai Jiao Tong University*, Shanghai, China, 263861852@qq.com
Haoran Wei
*Department of ECE*, *University of Texas at Dallas*, Richardson, USA, haoran.wei@utdallas.edu

**Human action detection is a hot topic, which is widely used in video surveillance, human–machine interface, healthcare monitoring, gaming, dancing training and musical instrument teaching. As inertial sensors are low cost, portable, and having no operating space, it is suitable to detect human action. In real-world applications, actions that are of interest appear among actions of non-interest without pauses in between. Recognizing and detecting actions of interests from continuous action streams is more challenging and useful for real applications. Based on inertial sensor and C-MHAD smart TV gesture recognition dataset, this paper utilized different inertial sensor feature formats, then compared the performance with different deep neural network structures according to these feature formats. Experiment results show the best performance was achieved by image based inertial feature with convolution neural network, which got 51.1% F1 score.**

## 1  INTRODUCTION

With the rapid growing of motion sensing game business, like Switch and Kinect devices, human action need to be detected to perform these games [1, 2, 3]. From these devices, inertial data including acceleration and angular velocity signals are collected simultaneously. With deep learning approaches, fast and accurate human action detection can be achieved by utilizing these inertial data. Human action detection is also widely used in video surveillance [4, 5], human–machine interface [6, 7], healthcare monitoring [8, 9], dancing training and musical instrument teaching [10, 11, 12].

Many devices can be used to detect human action, like video camera [13, 14, 15, 16, 17, 18], depth camera [19, 20], and inertial sensors [21]. Cameras have limited areas of observation and cannot detect actions beyond their field of view. Cameras are also very sensitive to light changing. As inertial sensors are low cost, portable, and having no operating space, it is suitable to detect human action.

Most of the previous research papers assume the actions of interest are already segmented, in other words, they assume the starting point and ending point are already known [22, 23, 24]. While in real-world applications, actions of interest appear among actions of non-interest without pauses in between. Recognizing and detecting actions of interests from continuous action streams is more challenging and useful for real-world applications. Papers [25, 26] performed continuous human action detection using a small dataset. Recently, paper [27] release a larger scale continuous human action detection dataset called C-MHAD (Continuous Multimodal Human Action Dataset).

With C-MHAD smart TV gesture recognition dataset, this paper utilized different inertial sensor feature formats, then compared the performance with different deep neural network structures according to these feature formats. The major contribution of this paper includes:
1)    Propose a deep learning based human action recognition approach using inertial data;
2)    Compare different feature formats of inertial data;
3)    Compare the performance of various deep learning models for continuous human action detection.

The rest of the paper is organized as follows: Section 2 cover the inertial data and different inertial feature formats that have been used for human action recognition. Then, the corresponding deep learning classification models are described in Section 3. Section 4 explain the experiment setup and present the experiment results. Finally, the paper is concluded in Section 5.

## 2 INERTIAL DATA AND INERTIAL FEATURE FORMATS

### 2.1 C-MHAD

Most of previous datasets assume the staring and ending point of an action is known, like Berkeley MHAD and UTDMHAD. While a more common application aims to detect actions of interest from continuous action stream. C-MHAD provide continuous action streams for human action recognition, consisting of video, depth and inertial data.

C-MHAD collected 240 continuous data streams from 12 subjects, 120 continuous data streams are used for smart TV gestures application, the rest 120 streams are used for transition movements application. As TV gestures application is closer to daily life, this paper will conduct experiment for this application.

For TV gestures application, 12 subjects (10 male and 2 female subjects) involved in data collection. Every subject collected 10 continuous data streams, each stream last for 2 minutes. Five actions are treated as actions of interest, they are swipe left, swipe right, wave, draw circle clockwise, and draw circle counterclockwise. Subject can play any other actions of non-interest in between. More details about this dataset are shown on [27].

### 2.2 Inertial Feature Formats

Inertial data consisting of 3-axis accelerations and 3-axis angular velocities. More specifically, the inertial data have x-axis acceleration, y-axis acceleration, z-axis acceleration, x-axis angular velocity, y-axis angular velocity and z-axis angular velocity.

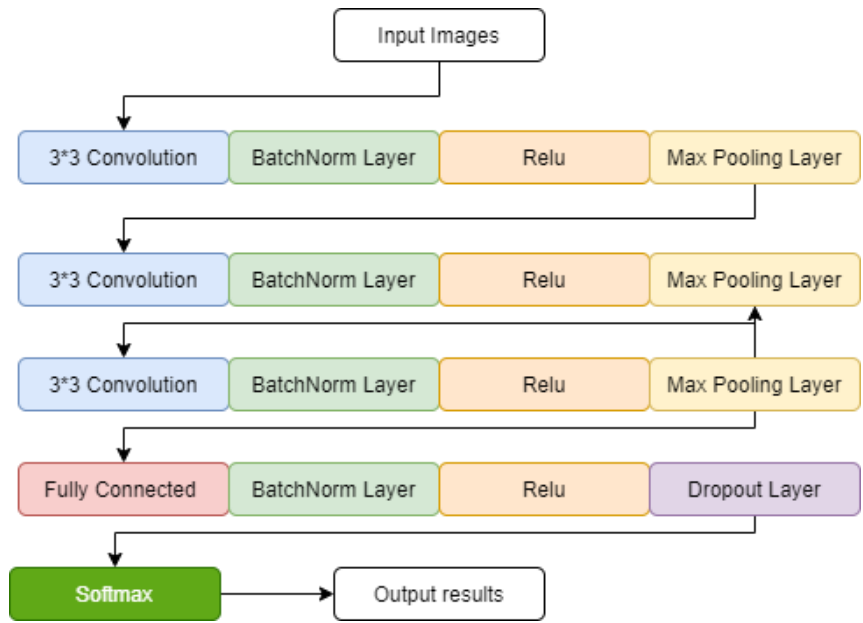

Figure 1. Example of the convolution neural network model.

After getting these 6 signals, normalization along each signal is performed. Then, the average acceleration a(t) and average angular velocity g(t) are calculated as follow:

$$a(t) = \sqrt{a_x(t)^2 + a_y(t)^2 + a_z(t)^2} \qquad (1)$$

$$g(t) = \sqrt{g_x(t)^2 + g_y(t)^2 + g_z(t)^2} \qquad (2)$$

After getting the average acceleration and average angular velocity at time t, there are 8 inertial signals at time t.

Then the max value (Max) and min value (Min) along each axis are selected out. Based on these values, each axis is normalized to range 0 to 1. For example, the average acceleration a(t) at time t is normalized as follow:

$$a(t) = (a(t) - Min)/(Max - Min) \qquad (3)$$

After the above processing, there is a 150 (frames) * 8 (dimension) image in three seconds duration. Then a median filter of size 3 is used on that matrix, making it a 50*8 input image.

This paper compared two different kind of inertial feature formats. One of the formats is descried above, using that image as a input to convolution neural network, then getting the classification results. The second format is based on the first format, it will calculate the mean and variance along each dimension from the 50*8 image. Then the mean and variance value from each dimension are concatenated together, making it a 16-dimension feature vector. After that this feature vector is feed into a fully connected neural network, and then getting the classification results from last layer.

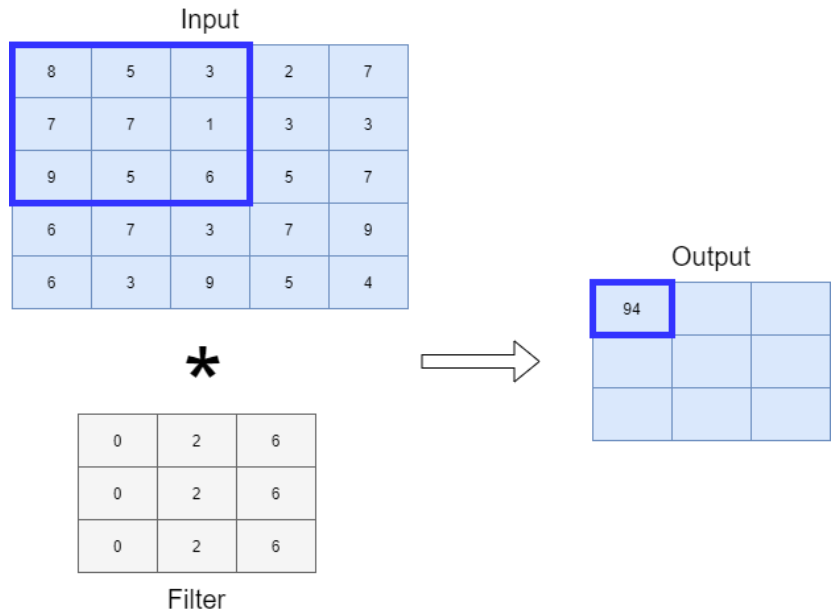

Figure 2. Example of convolution layer.

## 2.3 Classification Models for Continuous Action Detection

Machine learning [28, 29, 30, 31, 32]and deep learning [33, 34, 35, 36, 37, 38, 39, 40] are useful tools to deal with classification problems [41, 42, 43, 44]. After getting the inertial data, deep neural network models are utilized to get the action detection results. The convolution neural network model utilized in this paper is shown on Figure 1. There are 3 convolution layers in this convolution neural network model, with 16, 32 and 64 convolution filters respectively. Figure 2 gives an example of convolution layer. Then another fully connected layer and softmax function are utilized to get the final output results.

To be consistent with convolution neural network model, the fully connected deep neural network model replace the convolution layer with fully connected layer, making fully connected model have 4 fully connected layers in total. Each fully connected layer has 128 units in this paper.

BatchNorm layer performs mean and variance normalization for last layer input of each batch. Relu activation function is a non-linear function, it can reduce the size of parameter and solve the over-fitting problem. The equation of Relu function is as follow:

$$x = \begin{cases} 0, x < 0 \\ x, x \geq 0 \end{cases} \quad (4)$$

Pooling layer can compress the data dimensions and accelerate the computation. Max-pooling selects the maximum value from its receptive field. Figure 3 gives an example of max-pooling layer. Fully connected layer connects adjacent two layers, Figure 4 gives an example of fully connected layer. Dropout operation can reduce the over-fitting by ignoring some of the connections. Figure 5 gives an example of fully connected layer with dropout. Softmax function is commonly used in the last layer for classification tasks, it can be expressed as:

$$S_i = \frac{e^{f_i}}{\sum_j e^{f_i}} \quad (5)$$

in this equation, $f_i$ indicates the input value of each class. $S_i$ indicates the output value of each class. A bigger $S_i$ means the higher probability for belong to the $i_{th}$ class. The sum of all the $S_i$ equals to 1.

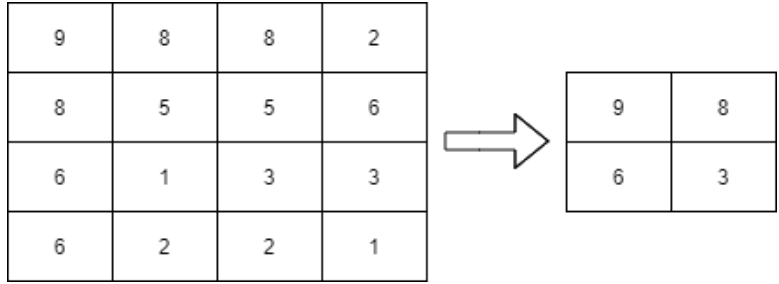

Figure 3. Example of max-pooling layer.

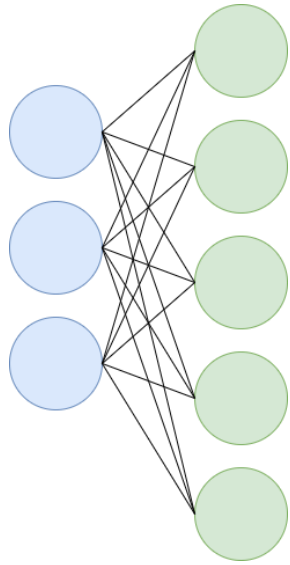

Figure 4. Example of a fully connected layer.

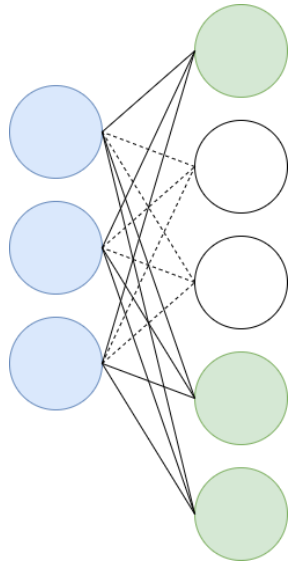

Figure 5. Example of a fully connected layer with dropout.

3    EXPERIMENTS SETUP AND RESULTS

For every subject from the TV gestures application dataset, streams number 1 to 9 are utilized to train the deep neural network, stream number 10 is utilized to test the performance. The training stage can divide into two phases, the first phase detect action of interest of all the actions (detection phase). The second phase classify the detected actions of interest into five target classes (classification phase). Both of the two phases can have errors, in the detection phase, action of non-interest can be detected as action of interest, action of interest can also been ignored. In the classification phase, the five actions of interest can be misclassified.

Table 1: Performance on Phase One

| Model | Predict Positive | Predict Negative |
|---|---|---|
| **Actual Positive** | TP | FN |
| **Actual Negative** | FP | TN |

Phase one is a two classes classification problem. There are four results for two this problem, they are TP (true positive), FN (false negative), FP (false positive) and TN (true negative). The relation between these four results are shown in Table 1 The TP in this application indicate corrected detected actions of interest. Based on the numbers (N) of the above four results, the precision, recall and F1 can be calculated as follow:

$$Recall = \frac{N_{TP}}{N_{TP}+N_{FN}} \quad (6)$$

$$Precision = \frac{N_{TP}}{N_{TP}+N_{FP}} \quad (7)$$

$$F1 = \frac{2 \times Precesion \times Recall}{Precision+Recall} \quad (8)$$

The F1 score from phase one is shown in Table 2. Phase two classified the results getting from phase one. After phase two, the error will be accumulated, so the new TP in phase two indicate action of interest which is correct detected and correct classified. Combining errors in two phase, the final F1 score is shown in Table 3.

Table 2 indicates image based inertial features with convolution neural network have better performance than vector based inertial features with fully connected neural network in the detection phase. Table 3 indicate the vector based inertial features have poor classification performance, so the convolution neural network with image based inertial feature format is a better solution for this continuous TV gesture detection application.

Table 2: Performance on Phase One

| Model | Precision | Recall | F1 |
|---|---|---|---|
| Convolution Neural Network | 48.8% | 62.0% | 54.6% |
| Fully Connected Neural Network | 36.8% | 43.0% | 39.6% |

Table 3: Performance on Phase Two

| Model | Precision | Recall | F1 |
|---|---|---|---|
| Convolution Neural Network | 45.7% | 58.0% | 51.1% |
| Fully Connected Neural Network | 5.1% | 6.0% | 5.5% |

## 4  CONCLUSION

Human action detection widely used in various applications. As inertial sensors are low cost, portable, and having no operating space, inertial sensors become a common device to detect human action. In real-world applications, continuous human action detection is more challenging and useful. Based on inertial sensor and C-MHAD smart TV gesture recognition dataset, this paper utilized different inertial sensor feature formats, then compared the performance with different deep neural network structures according to these feature formats. Experiment results show the best performance was achieved by image based inertial feature with convolution neural network.

REFERENCES

[1] V. Bloom, D. Makris, and V. Argyriou, "G3d: A gaming action dataset and real time action recognition evaluation framework," in *2012 IEEE Computer Society Conference on Computer Vision and Pattern Recognition Workshops*. IEEE, 2012, pp. 7–12.

[2] Y. Wang, T. Yu, L. Shi, and Z. Li, "Using human body gestures as inputs for gaming via depth analysis," in *2008 IEEE International Conference on Multimedia and Expo*. IEEE, 2008, pp. 993–996.

[3] S. K. Yadav, K. Tiwari, H. M. Pandey, and S. A. Akbar, "A review of multimodal human activity recognition with special emphasis on classification, applications, challenges and future directions," *Knowledge-Based Systems*, p. 106970, 2021.

[4] H. Wei and N. Kehtarnavaz, "Semi-supervised faster rcnn-based person detection and load classification for far field video surveillance," *Machine Learning and Knowledge Extraction*, vol. 1, no. 3, pp. 756–767, 2019.

[5] H. Wei, M. Laszewski, and N. Kehtarnavaz, "Deep learning based person detection and classification for far field video surveillance," in *2018 IEEE 13th Dallas Circuits and Systems Conference (DCAS)*. IEEE, 2018, pp. 1–4.

[6] L. L. Presti and M. La Cascia, "3d skeleton-based human action classification: A survey," *Pattern Recognition*, vol. 53, pp. 130–147, 2016.

[7] X. Chu, W. Ouyang, H. Li, and X. Wang, "Structured feature learning for pose estimation," in *Proceedings of the IEEE Conference on Computer Vision and Pattern Recognition*, 2016, pp. 4715–4723.

[8] M. G. Amin, Y. D. Zhang, F. Ahmad, and K. D. Ho, "Radar signal processing for elderly fall detection: The future for in-home monitoring," *IEEE Signal Processing Magazine*, vol. 33, no. 2, pp. 71–80, 2016.

[9] S. Chernbumroong, S. Cang, A. Atkins, and H. Yu, "Elderly activities recognition and classification for applications in assisted living," *Expert Systems with Applications*, vol. 40, no. 5, pp. 1662–1674, 2013.

[10] Y. Zhiqiang, "Study of physical education teaching based on kinect," *Automation & Instrumentation*, p. 10, 2016.

[11] J. Wang, T. Liu, and X. Wang, "Human hand gesture recognition with convolutional neural networks for k12 double-teachers instruction mode classroom," *Infrared Physics & Technology*, vol. 111, p. 103464, 2020.

[12] S. K. Yadav, A. Singh, A. Gupta, and J. L. Raheja, "Real-time yoga recognition using deep learning," *Neural Computing and Applications*, vol. 31, no. 12, pp. 9349–9361, 2019.

[13] S. Majumder and N. Kehtarnavaz, "A review of real-time human action recognition involving vision sensing," in *Real-Time Image Processing and Deep Learning 2021*, vol. 11736. International Society for Optics and Photonics, 2021, p. 117360A.

[14] H. Zhu, H. Wei, B. Li, X. Yuan, and N. Kehtarnavaz, "A review of video object detection: Datasets, metrics and methods," *Applied Sciences*, vol. 10, no. 21, p. 7834, 2020.

[15] Y. Liu, K. Wang, G. Li, and L. Lin, "Semantics-aware adaptive knowledge distillation for sensor-to-vision action recognition," *IEEE Transactions on Image Processing*, 2021.

[16] J. Chen, J. Kong, H. Sun, H. Xu, X. Liu, Y. Lu, and C. Zheng, "Spatiotemporal interaction residual networks with pseudo3d for video action recognition," *Sensors*, vol. 20, no. 11, p. 3126, 2020.

[17] H. Zhu, H. Wei, B. Li, X. Yuan, and N. Kehtarnavaz, "Real-time moving object detection in high-resolution video sensing," *Sensors*, vol. 20, no. 12, p. 3591, 2020.

[18] J.-K. Tsai, C.-C. Hsu, W.-Y. Wang, and S.-K. Huang, "Deep learning-based real-time multiple-person action recognition system," *Sensors*, vol. 20, no. 17, p. 4758, 2020.

[19] Z. Ahmad and N. Khan, "Cnn based multistage gated average fusion (mgaf) for human action recognition using depth and inertial sensors," *IEEE Sensors Journal*, 2020.

[20] J. Lee and H. Jung, "Tuhad: Taekwondo unit technique human action dataset with key frame-based cnn action recognition," *Sensors*, vol. 20, no. 17, p. 4871, 2020.

[21] Z. Ahmad and N. Khan, "Inertial sensor data to image encoding for human action recognition," *IEEE Sensors Journal*, vol. 21, no. 9, pp. 10978–10988, 2021.

[22] C. Chen, R. Jafari, and N. Kehtarnavaz, "Utd-mhad: A multimodal dataset for human action recognition utilizing a depth camera and a wearable inertial sensor," in *2015 IEEE International conference on image processing (ICIP)*. IEEE, 2015, pp. 168–172.

[23] H. Wei, R. Jafari, and N. Kehtarnavaz, "Fusion of video and inertial sensing for deep learning–based human action recognition," *Sensors*, vol. 19, no. 17, p. 3680, 2019.

[24] C. Chen, R. Jafari, and N. Kehtarnavaz, "Improving human action recognition using fusion of depth camera and inertial sensors," *IEEE Transactions on HumanMachine Systems*, vol. 45, no. 1, pp. 51–61, 2014.

[25] N. Dawar and N. Kehtarnavaz, "Continuous detection and recognition of actions of interest among actions of non-interest using a depth camera," in *2017 IEEE International Conference on Image Processing (ICIP)*. IEEE, 2017, pp. 4227–4231.

[26] H. Wei and N. Kehtarnavaz, "Simultaneous utilization of inertial and video sensing for action detection and recognition in continuous action streams," *IEEE Sensors Journal*, vol. 20, no. 11, pp. 6055–6063, 2020.

[27] H. Wei, P. Chopada, and N. Kehtarnavaz, "C-mhad: Continuous multimodal human action dataset of simultaneous video and inertial sensing," *Sensors*, vol. 20, no. 10, p. 2905, 2020.

[28] C. Li, L. Tian. "Association between resting-state coactivation in the parieto-frontal network and intelligence during late childhood and adolescence," *American Journal of Neuroradiology*. 2014, 35(6), pp:1150-1156.

[29] Z. Zhang, C. Sun, C. Li, M. Sun. "Vibration based bridge scour evaluation: A data-driven method using support vector machines," *Structural monitoring and maintenance*. 2019, 6(2), pp:125-145.

[30] H. Wei, Y. Long, H. Mao. "Improvements on self-adaptive voice activity detector for telephone data," *International Journal of Speech Technology.* 2016, 19(3), pp:623-630.

[31] M. Sun, C. Li, H. Zha. "Inferring private demographics of new users in recommender systems," *In Proceedings of the 20th ACM International Conference on Modelling, Analysis and Simulation of Wireless and Mobile Systems,* 2017, pp. 237-244.

[32] Y. Zhang, Y. Long, X. Shen, H. Wei, M. Yang, H. Ye, H. Mao. "Articulatory movement features for short-duration text-dependent speaker verification. International Journal of Speech Technology," 2017 Dec;20(4), pp：753-759.

[33] Y. Long, R. He. "The SHNU System for the CHiME-5 Challenge," *Proc. CHiME 2018 Workshop on Speech Processing in Everyday Environments.* 2018, pp: 64-66.

[34] H. Wei, N. Kehtarnavaz. "Determining number of speakers from single microphone speech signals by multi-label convolutional neural network," *In IECON 2018-44th Annual Conference of the IEEE Industrial Electronics Society,* 2018, pp. 2706-2710.

[35] R. Su, F. Tao, X. Liu, H. Wei, X. Mei, Z. Duan, L. Yuan, J. Liu, Y. Xie. "Themes Informed Audio-visual Correspondence Learning," *arXiv preprint* arXiv:2009.06573, 2020.

[36] K. Killamsetty, C. Li, C. Zhao, R. Iyer, F. Chen. "A Reweighted Meta Learning Framework for Robust Few Shot Learning," *arXiv preprint* arXiv:2011.06782, 2020.

[37] H. Wei, F. Tao, R. Su, S. Yang, J. Liu. "Ensemble Chinese End-to-End Spoken Language Understanding for Abnormal Event Detection from audio stream," *arXiv preprint* arXiv:2010.09235. 2020.

[38] L. Peng, A. Jiang, H. Wei, B. Liu, M. Wang. "Ensemble single image deraining network via progressive structural boosting constraints," *Signal Processing: Image Communication.* 2021, p.116460.

[39] Y. Shi, J. Zhou, Y. Long, Y. Li, H. Mao. "Addressing Text-Dependent Speaker Verification Using Singing Speech," *Applied Sciences*, 2019, 9(13), 2636.

[40] Y. Long, Y. Li, B. Zhang. "Offline to online speaker adaptation for real-time deep neural network based LVCSR systems," *Multimedia Tools and Applications*,2018, 77(21), pp:28101-28119.

[41] X. Gong, Y. Zhu, H. Zhu, H. Wei. "ChMusic: A Traditional Chinese Music Dataset for Evaluation of Instrument Recognition," *arXiv preprint* arXiv:2108.08470. 2021.

[42] H. Wei, A. Sehgal, N. Kehtarnavaz. "A deep learning-based smartphone app for real-time detection of retinal abnormalities in fundus images," *In Real-Time Image Processing and Deep Learning 2019*, 2019, Vol. 10996, p. 1099602).

[43] C. Zhao, C. Li, J. Li, F. Chen. "Fair meta-learning for few-shot classification," *In 2020 IEEE International Conference on Knowledge Graph (ICKG),* 2020, pp. 275-282.

[44] Y. Long, S. Wei, Q. Zhang, C. Yang. "Large-Scale Semi-Supervised Training in Deep Learning Acoustic Model for ASR," *IEEE Access*, 2019, (7), pp:133615-133627.